\documentclass[letterpaper,12point,epsfig]{article}
\usepackage{epsfig}

\begin{document}
\begin{center}
\huge{The Electrostatic Gravimeter:}

\huge{An Alternate Way of 
Measuring Gravitational Acceleration}
\end{center}
\vspace{.2 cm}
\begin{center}
\large{David O. Kashinski}

\small{Department of Physical Sciences, Kutztown University, Kutztown, 
Pennsylvania 19530}

\vspace {.15 cm}

\large{Paul V. Quinn Sr.}

\small{Department of Physical Sciences, Kutztown University, Kutztown, 
Pennsylvania 19530} 

\end{center}

\begin{abstract}
\noindent In the past, the gravitational acceleration on the surface of the Earth, $g$ has been measured in many 
ways.  Various methods include the use of a pendulum as well as models involving the use of a mass on a spring. 
We have designed a new method incorporating the mass-spring model with a capacitor, and a capacitance meter.  
This capacitor model still uses a hanging mass on a spring, but alters the method of determining the change in 
position of the mass attached to the spring caused by the gravitational acceleration.  We relate the 
change in position of mass to the capacitance across the two parallel plates.  By relating this capacitance 
directly to the gravitational acceleration, a new method of measuring $g$ is obtained.

\end{abstract}
\vspace{.1 cm}

\normalfont

\section{Introduction}

Gravity is defined as the force of attraction exerted between masses. [1] Here 
on Earth, it is commonly described as the force causing objects to fall toward the 
center of the Earth. In mechanics, it is a common practice for students to calculate the value of
the acceleration due to the Earth's gravitational field acting on other masses.  
In the classroom, this gravitational acceleration is commonly referred to as the constant 
$g$, with an accepted value of $9.81 m/s^2$.  In reality, the value of $g$ is 
not constant, but rather, changes over the surface of the Earth.  The variations in the Earth's radius are 
responsible for the variation of $g$.  Using Newton's theory of gravity, we 
derive an exact expression for the gravitational acceleration,

$$g = \frac{GM_E}{r^2},\eqno {(1)}$$

\noindent where $M_E$ is the mass of the Earth, $G$ is the gravitational constant, and $r$ is the 
distance to the center of the earth.[1]  Depending on where you are on the surface of the Earth, the value of 
$g$ will fluctuate.  The value for $g = 9.81 m/s^2$ used in most introductory physics courses is 
actually the value of $g$ at sea level.  

In the field of geology, the value of $g$ can be used to determine many characteristics of the Earth 
such as rock density, porosity, layer composition, and locations of underground aquafers, caverns, or air 
pockets.  By monitoring the slight changes of the value of $g$, geologists can extrapolate information about the 
structure of the Earth's crust below the surface.  All methods of extrapolating data from the measurements of 
$g$ can be derived from Newton's law of gravity[3],

$$F_g = \frac{GM_Em}{r^2}. \eqno {(2)}$$

\noindent Specialized methods of measurement using various types of measuring devices called gravimeters have been 
designed by geophysicists to easily obtain the value or the changes in the value of $g$ at various positions 
on the Earth's surface[2].  
	
\section{Current Techniques of Measuring Gravity}

One method used for obtaining  a value for $g$ is the "free fall" method. This involves dropping an object of 
mass $m$ from a distance $y$ above the Earth's surface.  We can then measure the time $t$ in which it 
takes the mass to traverse the distance.  Using the following projectile motion equation, 

$$\Delta y = \frac{1}{2}gt^2,\eqno {(3)}$$

\noindent we can use the time $t$ it takes to fall and the distance $y$ that it falls to calculate the value of 
the gravitational acceleration $g$.[3]  This method is frequently used in most high school and introductory 
college physics courses.  However, the value obtained with this simple calculation is not precise enough to 
significantly change when being measured from various locations on the surface of the Earth.  To obtain the 
information needed by geologists, a much more accurate method has to be used.  
	 
A second more accurate method procedure for obtaining the value $g$ requires the use of a simple pendulum.  In 
an introductory physics or mechanics class, students learn that the period of a pendulum can be expressed 
as

$$T = \frac{1}{2 \pi}\sqrt{\frac{L}{g}},\eqno {(4)}$$
   
\noindent where $T$ is the period of the pendulum and $L$ is the length of the pendulum.[4]  The pendulum can 
be used to determine a much more precise value for the gravitational acceleration.  There are some geophysicists 
that may use a  duoble pendulum gravimeter to conduct preliminary measurements of $g$.  However, the changes that 
occur as you move across the surface of the Earth are still not accurate enough to obtain some of the more 
detailed information about the Earth's crust.  Plus, the pendulum apparatus is cumbersome to move around, and the 
measurement process itself is fairly time consuming.[5]    
	 	
The method used most commonly by geophysicists interested in obtaining an 
accurate measurement of the Earth's gravitational field, requires the analysis of 
a mass-spring system.  A mass hanging from a stationary spring is under the influence
of the following two forces: the force of gravity pulling it down and the force of the
spring pulling it up.  These two forces are equal and opposite, causing the block to be motionless while in a 
state of static equilibrium.  One can obtain the value of the gravitational acceleration by setting
the two forces equal to one another and solving for $g$ as shown below,

$$mg = k y\eqno {(5)}$$
$$g = \frac{k}{m}y,\eqno {(6)}$$

\noindent where $k$ is the spring constant and $y$ is the distance the string has stretched from its
equilibrium position without the mass as shown in Fig. 1.  If the mass and the spring constant are 
known, you can calculate the gravitational constant by measuring the distance the spring is stretched 
from its initial equilibrium position.  This is the device that is commonly used by many geophysicists
to obtain measurements of the gravitational acceleration.[6]  However, such a device, called a gravimeter, 
can cost anywhere from \$15,000 to \$50,000 or more.  The most difficult aspect in the design of the gravimeter 
is determining the changes in the stretch of the spring with slight changes in $g$ as you survey different 
locations on the surface of the Earth.  Changes in $y$ are to small to detect with the human eye.  These small 
minute changes are detected via a complicated processe using a LASER within the apparatus itself.  The bulk of 
the cost of a gravimeter comes from the various apparati designed to measure the slight changes in $y$.[7,8]   
 
The purpose of this paper is to propose an easier method of measuring the slight changes in the $y$ used in the 
mass-spring gravimeter while providing close to the same accuracy when determining $g$. This new gravimeter also 
has educational value in both physics and geology.  We will show that with a simple application of 
elementary electrostatics, the same measurements are possible with a lower cost and a less complicated 
apparatus.  For geologists, this may provide an easy-to-build, inexpensive gravimeter yielding fairly 
accurate measurements of $g$.  For for college faculty, it will provide an excellent way to demonstrate 
changes in $g$ using simple concepts of physics.  Our design is called the electrostatic gravimeter, and it 
uses a capacitor as the primary component in the device for measuring changes in $y$.   

\section{The Capacitor}

Before we discuss the detailed design of the electrostatic gravimeter, we will first review some of the 
electrostatic theory used to describe capacitors.  There are many types of capacitors used in physics, 
specifically in the field of electronics.  In the electrostatic gravimeter, a parallel plate capacitor is 
used in the design.  A parallel plate capacitor is two identical metallic plates with surface area $A$, 
separated by some distance $d$.  A parallel plate capacitor connected to a DC power supply as depicted in the 
in the schematic of Fig. 2, has many useful properties, making it a key component in many electronic devices.[9]

Capacitance is defined as the ratio of the magnitude of the total charge on one of the metal plates to the 
potential difference between them [9],

$$C = \frac{q_{tot}}{V}. \eqno {(7)}$$

\noindent Using our knowledge of electrostatics, we can use Gauss's Law,

$$ \oint {\bf{E}} \cdot {\bf{da}} = \frac{q_{enc}}{\epsilon_o},\eqno {(8)}$$

\noindent to derive an expression for the electric field $E$, between the two plates, where $q_{enc}$ is just 
$q_{tot}$ on one plate.  Solving for $q_{tot}$, we obtain the following expression:

$$q_{tot} = \epsilon_o E A,\eqno {(9)}$$

\noindent where $E$ is the magnitude of the electric field, $q_{tot}$ is the total charge on one of the plates, 
and $\epsilon_o$ is the permittivity of free space.  Solving this expression for the field, we get

$$E=\frac{q_{tot}}{\epsilon_o A}. \eqno {(10)}$$

To obtain an expression for the capacitance, the potential difference is calculated using

$$V=\int_+^- {\bf{E}} \cdot {\bf{ds}}. \eqno {(11)}$$

\noindent We assume the field is uniform and that the plates are large enough and close enough together such that 
no fringing of the electric field occurs at the ends of the plates.  In Eq.(11) we are integrating from the 
positively charged plate to the negatively charged plate.  Changing the limits of integration and evaluating 
the scalar product, the integral becomes

$$V=\int_0^d E ds. \eqno {(12)}$$

\noindent Using the uniformity of the field, we integate Eq.(12) and get 

$$V = E d.\eqno {(13)}$$

\noindent We now obtain the expression for the capacitance of a parallel plate 
capacitor in free space[10],  

$$C = \frac{\epsilon_o A}{d}. \eqno {(14)}$$

The capacitance derived in Eq.(14) is for two metallic plates separated by a vacuum.  However, in most 
situations, capacitors are not used in a vacuum.  There is usually a material called a dielectric in between 
the plates of the capacitor.  As the electric field interacts with the molecules of the dielectric, they line up 
with the field, producing their own electric field opposite the direction of the original.  This causes a lower 
value of the total electric field between the plates.  Since the value of the capacitance is inversely 
proportional to $E$, the value of $C$ will increase when a material is added between the plates of the capacitor.	
The capacitance with the dielectric, denoted as $C^\prime$, is proportional to the capacitance in a vacuum as 
shown below:

$$C^\prime = \kappa C,\eqno {(15)}$$

\noindent where $\kappa$ is the dielectric constant of proportionality. Since $C^\prime$ is always bigger than 
$C$, we can deduce that $\kappa > 1$.  The value of $\kappa$ will depend on the material used as the dielectric.   Combining 
Eq.(14) and (15), we obtain an expression for the capacitance of a parallel plate capacitor with a 
dielectric material between the plates [10],

$$C^\prime = \kappa \frac{\epsilon_o A}{d}.\eqno {(16)}$$

\noindent Often, the constants $\kappa$ and $\epsilon_o$ are combined into one constant called the permittivity 
constant, $\epsilon = \epsilon_o \kappa$.  Using the pemittivity constant, we can rewrite the parallel plate 
capacitance with a dielectric as
 
$$C^\prime = \epsilon\frac{A}{d}. \eqno {(17)}$$ 
 
\section{The Electrostatic Gravimeter}

Air as a dielectric allows us to obtain an expression for $g$, the acceleration do to gravity, in terms of the 
capacitance between two plates used in our new gravimeter.  Starting with the mass-spring system, we know that 
$y$, the distance the spring is pulled from its equilibrium position by the mass, is determined by the 
gravitational acceleration $g$.  By measuring $y$, one can essentially calculate the value of $g$ at a 
particular location on earth.  We can use a capacitor to accurately determine the distance $y$ and how it changes 
with your location on the Earth's surface. A sketch of the proposed apparatus is shown in Fig.3.
In the diagram, we have a mass-spring system made of a metallic conductor.  The mass-spring system is 
contained inside an insulated cylinder, surrounded by air at atmospheric pressure.  The spring is attached to 
the top of the cylinder.  A cylindrical mass, made of the same material as the spring, is permanently attached to 
the other end of the spring.  The mass is freely hanging, but is just slightly smaller in diameter, than the 
cylinder that contains it.  At the bottom of the cylinder is a second metal plate also made of the same material 
as the spring.  Both metallic plates will be connected to a capacitance meter, which has a small internal power 
supply used to apply voltage across the plates to assist in measuring the capacitance.  The plates themselves 
will be the capacitor in the system with air acting as the dielectric between them.  The the capacitance meter 
should be very precise in its output, to allow us to calculate $g$ as accurately as possible.  Because the 
distance between the two plates will change due to changes in the graviational acceleration gravity over various 
locations on the Earth's surface, the capacitance will change as measured by the meter attached to the two 
plates.  Therefore, by measuring the changes in the capacitance between the two plates, we can directly 
calculate variations in the Earth's gravitational field.  

To derive an expression for $g$, we will the diagram presented in Fig.3.  Our first step was to obtain an 
expression for $d$, the distance between the plates, in terms of the variables shown in Fig.3.  From the diagram, 
we find that

$$d=h-a-c,\eqno {(18)}$$

\noindent where $a$ is the thickness of the cylindrical mass, $c$ is the length of the spring in its stretched 
position, and $h$ is the height inside the cylinder.  As illustrated in Fig.1, the variable $c$ can be defined 
as

$$c=y + y_o,\eqno {(19)}$$

\noindent where $y_o$ is the length of the spring with no mass attached, and $y$ is the additional stretched 
length that occurs when the mass is added.  If $g=0$, then $y = 0$ and the value of $c$ would simply by the 
length of the spring.  Using Eq.(6), (18) and (19), we obtain the following expression for $d$:

$$d=h-a-y_o-\frac{mg}{k}.\eqno {(20)}$$

Using Eq.(14) and (20) we found that, 

$$C = \frac{\epsilon A}{d}=\frac{\epsilon A}{h-a-y_o-mg/k},\eqno {(21)}$$

\noindent where $\epsilon$ would account for any dielectric material, in this case air, placed between the two 
cylindrical plates.  Since we are working with a cylinder, the area of both plates will be circular in nature, 
making $A = \pi r^2$.  The shape of your plates could be square or rectangular, but for convenience, we chose to 
derive the result using a circular system.  Plugging in the value of $A$ and solving for $g$, we get

$$g = \frac{k}{m}(h-a -y_o-\frac{\epsilon \pi r^2}{C}).\eqno {(22)}$$

This gives us an expression for $g$ in terms of the capacitance $C$ between the cylindrical plates as well as 
the physical dimensions and properties of the system.

\section {Range of Gravimeter Measurements}

The electrostatic gravimeter gives us a new method of determining gravitational acceleration, simply by measuring 
the capacitance across the two parallel plates inside the apparatus.  The process for obtaining $C$ is simply a 
matter of reading a value given by a very precise meter.  You can then determine $g$ by plugging the appropriate 
values inot Eq.(22).  By looking at some realistic values, we can show that small changes in the gravitational 
acceleration produce changes in the capacitance that are easily measureable with a precise meter.  On the surface 
of the earth, gravitational acceleration ranges from approximately $9.78306 m/s^2$ to $9.83208m/s^2$.[11]  This 
is a total range of $\Delta g \approx 0.04902m/s^2$.  To obtain a range for the change in capacitance, we chose 
the following reasonable values for the dimensions of the apparatus and the ratio of the mass and the spring 
constant $h = .3 m$, $a = .0051 m$, $y_o =.285 m$, $r = 0.15 m$, $m/k = .001 kg N/m $, and 
$\epsilon = 8.8552215 \times 10^{-12}$.  The value of $\epsilon$ used here is the permeability constant of 
air.[12]  These values are all reasonable for the apparatus proposed in this paper.  The values may seem very 
exact, but an apparatus with these exact values can be easily manufactured by a skilled machinist.  Using these 
values, we find the range of $C$ to be approximately $5.353 \times 10^{-9}$ - $9.216 \times 10^{-9}F $.  To 
measure changes in gravitational acceleration on Earth, we need to detect changes in capacitance of at most 
$\Delta C \approx 3.863 \times 10^{-9} F$.  In the field of electronics, there are meters that can easily detect 
changes in capacitance on the order of $1.00 \times 10^{-14}$.  

Using Eq.(1), it can easily be shown that a change in height of approximately $10 m$ results in a change in the 
gravitational acceleration of $\Delta g \approx 3.00 \times 10^{-5} m/s$.  This corresponds to a change in 
capacitance of $\Delta C \approx 4.475 \times 10^{-12} F$.  From this result, we can infer that one should easily 
be able to detect the change in $g$ that occurs when moving between two floors of a building.  This would be an 
excellent educational demonstration when discussing gravity in an introductory level physics class.  When a 
gravimeter is used to take measurements by geophysicists, changes in $g$ usually range from approximately 
$1.00 \times 10^{-5}$ to $1 \times 10^{-6} m/s^2$.  According to Eq.(21), this reqires looking at changes in 
capacitance as small as $\Delta C \approx 1.00 \times 10^{-14} F$.  Once again, this is 
in range of what is detectable using modern electronics.  

\section{Conclusion}
Gravitational acceleration $g$, is a numerical constant used in all levels of physics, from high school to 
graduate research in physics.  The value of $g$ is also used by geophysicists to determine various properties of 
the Earth's crust.  For geophysicists, this value becomes extremely important for mining or oil exploration.  
Currently, methods to obtain these value require very expensive equipment combined with very tedious 
and complicated measurements.[6,11]  For a thorough and detailed geological survey covering large areas of land, 
the more complicated apparatus is probably more appropriate.  However, for geological surverys conducted by 
faculty with or without students, for small research projects, our new electrostatic gravimeter would be more 
than suitable.  This electrostatic gravimeter would be ideal for small surveys about the size of a football 
field.  It could be used to conduct less complicated studies such as the location of small faults or crust 
composition on a small scale.  Projects such as this couls be completed without spending tens of thousands of 
dollars on the more complex apparatus.  The cost of manufacturing our gravimeter, would clearly be much less, the 
majority of the cost being the meter used to measure the capacitance. 

The electrostatic gravimeter could also be used as a wonderful educational tool in the classroom.  First of all, 
it is an excellent demonstration of how the simple concepts learned in an introductory physics class can be 
utilized to create a device used for important research.  Many times in an introductory class, students find 
a lot of the concepts they learn to be abstract and not practical.  The electrostatic gravimeter provides an 
example of the practical use of these concepts.  One could actually demonstrate to students how the gravitational 
acceleration changes as you change the height of the apparatus.  This gives students a real picture of how gravity 
works here on Earth.  A geology professor might also be able to use the device in a classroom or laboratory 
setting to demonstrate changes in $g$ over various locations on the surface of the Earth.  Because the design of 
this gravimeter is fairly straightforward, its construction could make a wonderful project for an undergraduate 
student.  It would be a great experience for a physics major with machining skills and an interest in 
experimental physics.  The electrostatic gravimeter has the potential to be a useful educational tool for 
both geologists and physicists.

\newpage

\Large
\noindent \bf References
\normalsize
\normalfont
\vskip .4 true cm 

\noindent [1] Raymond A. Serway and John W. Jewett, Jr., \it{Physics for Scientists and Engineers}
\normalfont 6th Ed., Thomson Learning Inc., pp. 389–441(2004).
 
\vskip .2 true cm

\noindent [2] Jon P. Davidson, Walter E. Reed, and Paul M. Davis, \it{Exploring Earth}
\normalfont 2nd Ed., Prentice Hall, pp.135-142(2002).

\vskip .2 true cm

\noindent [3] David Halliday, Robert Resnick, and Jearl Walker, \it{Fundamentals of Physics} 
\normalfont 7th Ed., John Wiley \& Sons, Inc., pp. 24-27(2005).

\vskip .2 true cm

\noindent [4] Grant R. Fowles and George L. Cassiday, \it{Analytical Mechanics} \normalfont 6th Ed., 
Thomson Learning Inc., pp.79-80(1999). 

\vskip .2 true cm

\noindent [5] William Frederick Hoffmann, "{A Pendulum Gravimeter for Measurement of Periodic Annual 
Variations in the Gravitational Constant",\normalfont Thesis (Ph.D.), Princeton University, 
\it{Dissertation Abstracts International}, \normalfont Vol. 23-03, (1962).

\vskip .2 true cm

\noindent [6] Allen E. Mussett and M. Aftab Khan, \it{Looking Into The Earth},
\normalfont Cambridge University Press, pp.107-123 (2000).

\vskip .2 true cm

\noindent [7] Thomas M. Boyd, "Mass and Spring Measurements", 
http://www.mines.edu/

\noindent fs\_home/tboyd/GP311/MODULES/GRAV/NOTES/

\noindent spring.html(2002).

\vskip .2 true cm

\noindent [8] Micro-g Solutions, "FG-5 Absolute Gravimeter", 
http://www.microgsolutions.com/

\noindent fg5.htm, Micro-g Solutions Inc.(2002)

\vskip .2 true cm

\noindent [9] Paul A. Tippler and Gene Mosca, \it{Physics for Scientists and Engineers}
\normalfont 5th Ed., W.H. Freeman and Company, pp.752-775 (2004).

\vskip .2 true cm

\noindent [10] David J. Griffiths, \it{Introduction to Electrodynamics} \normalfont 3rd Ed., Prentice Hall, 
pp.103-106,179-196(1999)

\vskip .2 true cm
 
\noindent [11] Robert J. Lillie, \it{Whole Earth Geophysics},
\normalfont Prentice Hall, pp.223-275(1999).

\vskip .2 true cm

\noindent [12] David R. Lide, \it{CRC Handbook of Chemistry and Physics} \normalfont 83rd Ed.,  
CRC Press LLC, (20020
 
\vskip .2 true cm

\newpage

\Large
\noindent \bf Figure Captions
\normalsize
\normalfont
\vskip .4 true cm 

\noindent Figure 1:  A diagram of the mass-spring system used to determine the gravitational acceleration $g$.  
This figure illustrates the distance the spring stretches due to the mass attached at one end. 

\vskip .2 true cm

\noindent Figure 2:  A schematic o the parallel plate capacitor with a voltage source, which will be used by our 
electrostatic gravimeter.

\vskip .2 true cm

\noindent Figure 3:  A diagram of our proposed electrostatic gravimeter.  The cylinder bottom, the spring, and 
the hanging mass would all be constructed out of the same metallic material.  The rest of the cylinder would be 
an insulating material.  The capacitance meter would be attached to the top of the spring, and thebottom of 
the cylinder.  A cylindrical geometry was chosen out of convenience, but a rectangular system would work just as 
well.  

\newpage

\begin{figure}
\begin{center}
\epsfig{file=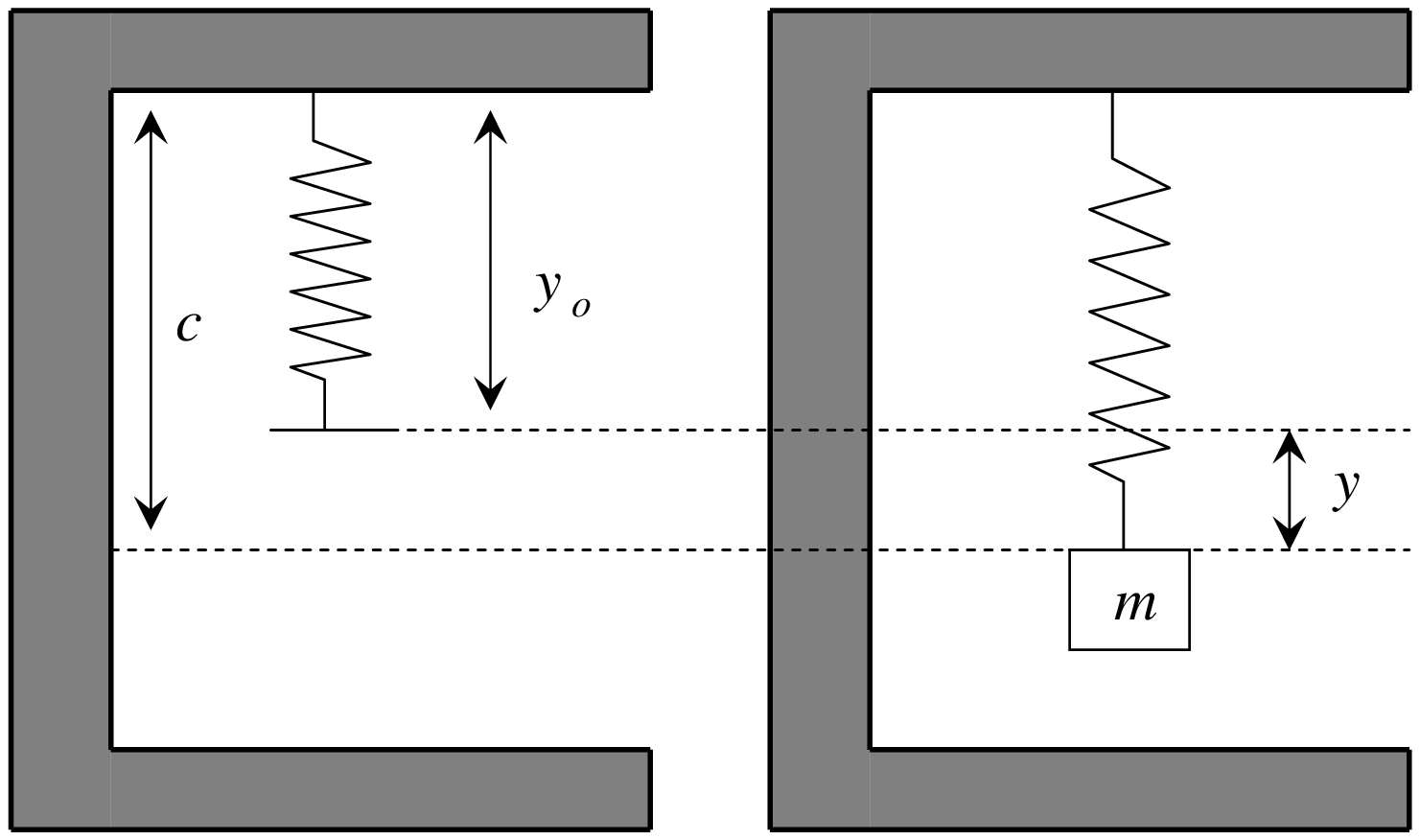,height=14cm,clip=}
\end{center}
\end{figure}
\newpage

\begin{figure}
\begin{center}
\epsfig{file=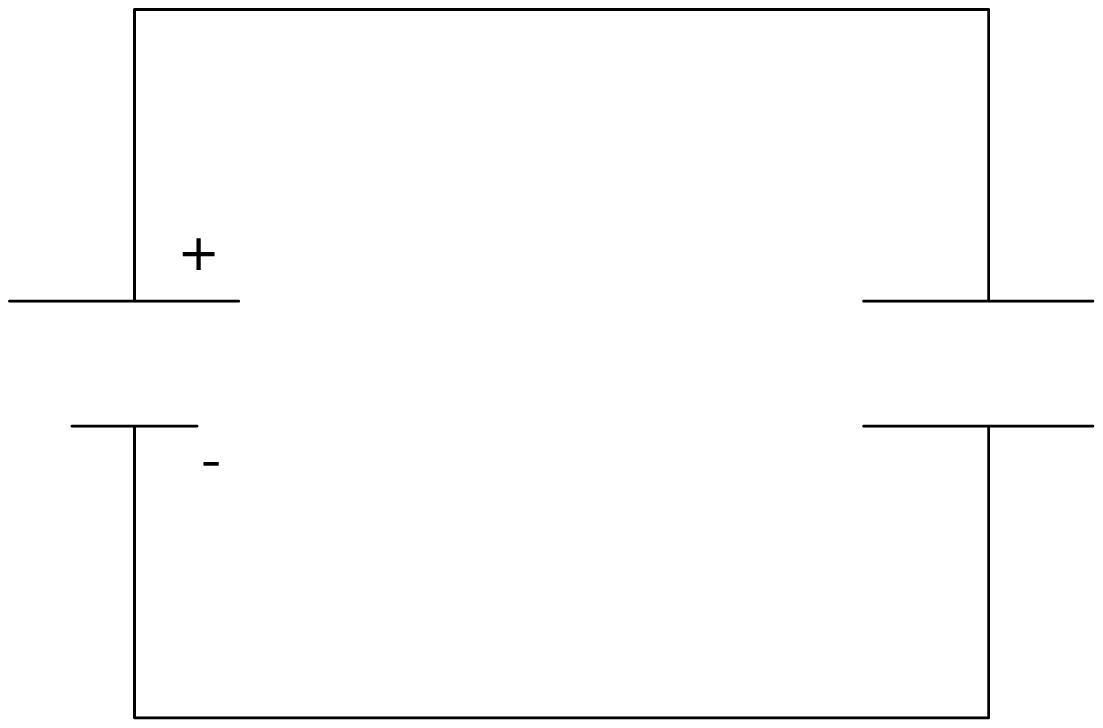,height=19cm,clip=}
\end{center}
\end{figure}

\newpage

\begin{figure}
\begin{center}
\epsfig{file=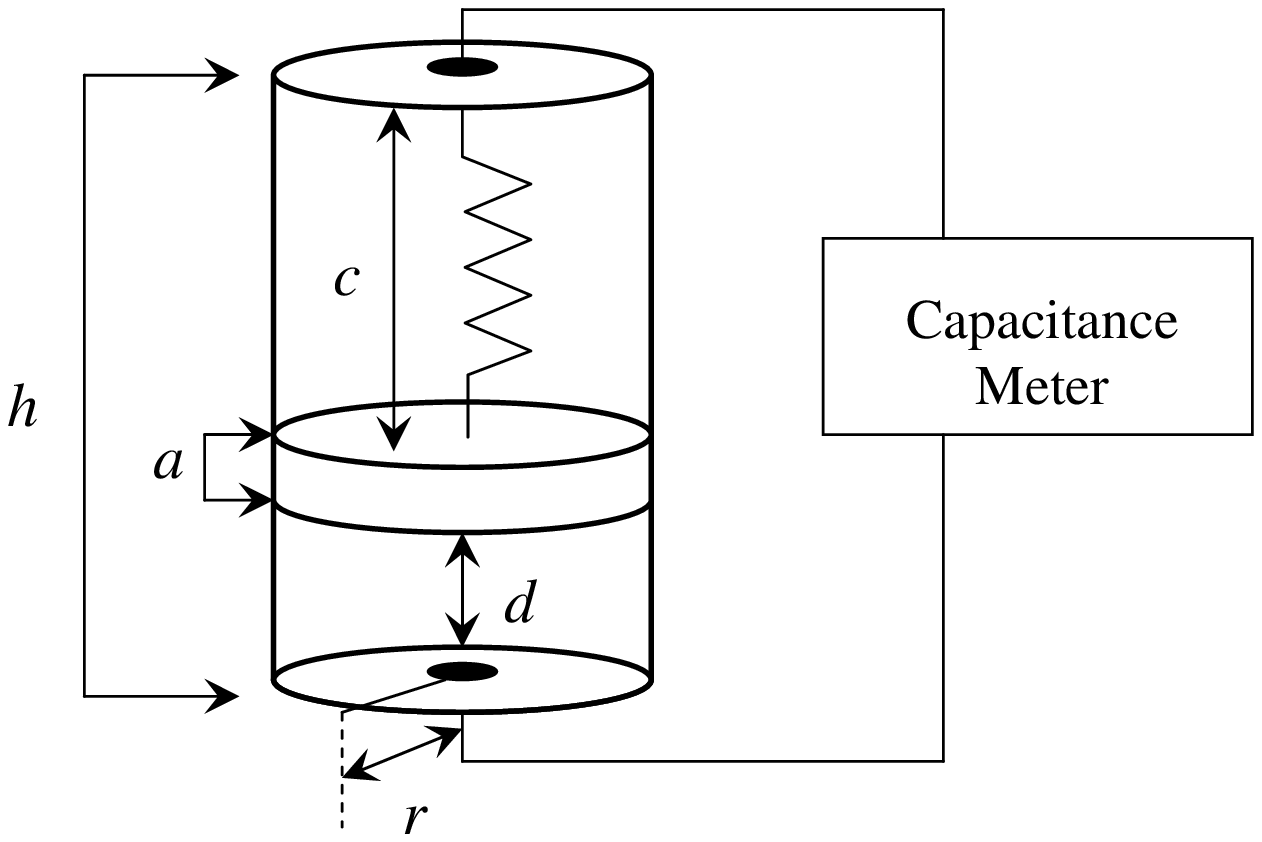,height=19.5cm,clip=}
\end{center}
\end{figure}

\end{document}